\documentclass[conference]{IEEEtran}
\IEEEoverridecommandlockouts
\usepackage{cite}
\usepackage{amsmath,amssymb,amsfonts}
\usepackage[english]{babel}
\usepackage{algorithmic}
\usepackage{graphicx}
\usepackage{hyperref}
\usepackage{cleveref}
\usepackage{mathrsfs}
\usepackage{textcomp}
\usepackage{xcolor}
\usepackage{multirow}
\usepackage{pgfplots}
\pgfplotsset{compat=1.11}
\def\BibTeX{{\rm B\kern-.05em{\sc i\kern-.025em b}\kern-.08em
    T\kern-.1667em\lower.7ex\hbox{E}\kern-.125emX}}
\graphicspath{ {./figure/} }
\begin{document}

\title{Multi-task recommendation system for scientific papers with high-way networks\ 
}

\author{\IEEEauthorblockN{1\textsuperscript{st} Aram Karimi}
\IEEEauthorblockA{\textit{University of Gothenburg} \\
aram.karimi@gu.se}
\and
\IEEEauthorblockN{2\textsuperscript{nd} Simon Dobnik}
\IEEEauthorblockA{\textit{University of Gothenburg} \\
simon.dobnik@gu.se
}
}

\maketitle

\begin{abstract}
Finding and selecting the most relevant scientific papers from a large number of papers written in a research community is one of the key challenges for researchers these days. As we know, much information around research interest for scholars and academicians belongs to papers they read. Analysis and extracting contextual features from these papers could help us to suggest the most related paper to them.
In this paper, we present a multi-task recommendation system (RS) that predicts a paper recommendation and generates its meta-data such as keywords.
The system is implemented as a three-stage deep neural network encoder that tries to
maps longer sequences of text to an embedding vector and learns simultaneously to predict the 
recommendation rate for a particular user and the paper's keywords.
The motivation behind this approach is that the paper's topics expressed as keywords are a useful predictor of preferences of researchers.
To achieve this goal, we use a system combination of RNNs, Highway and Convolutional Neural Networks to train end-to-end a context-aware collaborative matrix.
Our application uses Highway networks to train the system very deep, combine the benefits of RNN and CNN to find the most important factor and make latent representation.
Highway Networks allow us to enhance the traditional RNN and CNN pipeline by learning more sophisticated semantic structural representations.
Using this method we can also overcome the cold start problem and learn latent features over large sequences of text.


\end{abstract}


\section{Introduction}
Scientific research is a systematic approach to reporting advanced knowledge to the world. Many studies have looked at the growth rate of scientific publications and results shows that science has been progressed greatly over the past century. 
Finding and having access to publications among this huge number of articles, plays a key role for researchers. This 
motivates utilisation of 
this massive data and develop a practical system to 
suggest most relevant articles to researchers. 
Using a recommendation system (RS) could help to develop a systematic approach to filter out information and recommend papers that align to their research interests. \\
Generally, Recommendation system can be built 
using various method such as collaborative filtering (CF), a content-based and hybrid technique \cite{sharma2013survey}. CF methods produce user specific recommendations of items, based on patterns of ratings or usage without the need for 
external information about either items or users. In order to make 
recommendations, CF systems need to relate two fundamentally different entities: items and users. There are two main approaches to facilitate 
this in CF: the neighbourhood approach and latent factor models \cite{su2009survey}. Neighbourhood methods compute relationships between items or, alternatively, between users \cite{herlocker2002empirical}. Latent factor models, such as matrix factorisation, 
represent an 
approach by mapping items and users to a low-dimensional space, and compute a user-item preference score based on a function of two (i.e., user and item) latent vectors \cite{koren2015advances}. In content-based techniques, the system attempts to recommend items similar to those a user preferred in the past. In this case, the recommendations are based on 
the content of a given item, not on other users’ opinions as in the case of collaborative filtering. These techniques are vulnerable to over-fitting, but their advantage is, they don't need any data from other users side \cite{pazzani2007content}. 

Both the CF and the content-based approach have some weaknesses. In the CF this is the cold start issue and 
the primary drawback of content-based filtering systems is their tendency to over-specialise item selection  and predicting only very similar items to the previously seen items 
\cite{LU201512}. To overcome these issues, 
several approaches suggest using hybrid models that use both content and user-item patterns to make the best suggestions to users \cite{ghazanfar2010scalable}. 
Using a CF matrix and latent 
information from text is normally the best way to create hybrid models. This has potential to solve the cold start issue and also overcome the sparsity problem 
\cite{10.1145/1454008.1454053,wang2015collaborative}. Many older approaches such as topic modelling \cite{wang2011collaborative} and bag-of-words \cite{wang2015collaborative} are based on this strategy but their main 
shortcoming is they ignore the relation between words in the sequences of input text.

In recent years, a variety of deep learning models have been applied successfully 
in various applications 
that include text analysis, machine translation, image captioning, document summarisation, scientific paper recommendation systems and others \cite{pouyanfar2018survey}. Using RNN to extract latent 
information and representing text as embedding vectors \cite{cho2014learning} has been one of the main approach to modelling textual semantic information. \cite{wang2015collaborative,10.1145/1454008.1454053} 
apply deep learning on the task of recommendation systems for scientific papers. 
In \cite{10.1145/2959100.2959180} Trapit et al. 
implement an encoder using Gated Recurrent Neural Network to map text input to a latent factor vector. 
The model is trained end-to-end and has 
benefits such as taking into account the word order, the ability to make 
multi-task predictions of articles to users and tags for papers. However, for long sequences, RNNs, even when using a gating mechanism, 
identifying the relevant information if this can be found anywhere in a document.

In this paper we use a 3-stage deep neural network proposed in \cite{wen2016learning} for text representation. This includes an RNN, a highway layer and a CNN. This model tries to combine the benefits of RNNs and CNNs to extract the best information from long sequences of text. Combined with 
a highway layer 
the network is capable of learning structural representations and associating them 
with individual word representations. 
We combine our text mapping encoder with a CF technique \cite{10.1145/2959100.2959180} and model 
sequential language model as an input to supervised multi-task learning 
that predicts both the rate and the meta data. 
We also show how this approach can be used in the case of the cold start and user-item matrix sparsity.
We evaluate our model on the CiteULike dataset. 
Our results show that 
this approach improves the recommendation task for scientific papers. We compare our model with the GRU \cite{10.1145/1454008.1454053} and the embedding model \cite{wang2015collaborative} on both long and short sequences of text. 
We find that this model is very promising for the RSs task with a considerable 
improvement of performance compared to other models. Additionally, our analysis of how sequence length effects the the RCNNs with highway layers shows that this model learns good representations for long sequences of text.

\section{Background And Problem formulation}
\subsection{Scientific paper recommendation}
The number of academic studies is increasing every year and with the development of technology the interest in digital resources and access to information has increased over time. Researchers 
use and read exciting  articles to improve their studies and find new challenges in their scientific area. They spend 
a lot of time and struggle to find the relevant articles. 
The purpose of article recommendation system is to reduce this time and suggest to them the articles that aligns with their research interest and that they may not even be are of \cite{bulut2018paper}. Due to the large knowledge on the web and the need to filter it, recommendation systems (RS) have become 
vital tool \cite{DHANDA2016483}. Paper RSs aim to help researchers find the relevant papers by calculating and ranking publications 
and recommend a top N papers associated with a researcher's focus of interest\cite{bai2019scientific}. There 
several methods for scientific RS. \\
Most traditional research paper RSs rely on a keyword-based search technique \cite{sun2014leveraging,dai2018low}. The results of a keyword-based searching are not always suitable and the number of obtained items is relatively large. Researchers then have to filter the search results to get the items needed \cite{hassan2017personalized}.  Research paper RSs are mainly focused on 
the frequency of citations. Citation databases such as CiteSeerX\footnote{CiteSeerX \url{https://citeseerx.ist.psu.edu/index}} apply citation analysis in order to identify papers that are similar to a given paper \cite{west2016recommendation,zarrinkalam2013semcir,liu2015car}. Recently, making use of a matrix factorisation, a kind of CF technique, has become as the most efficient and accurate approach \cite{10.1145/1454008.1454053}. The approaches based on frequencies of citation and CF may face the cold start problem. 
Content-based methods, on the other hand, recommend items based on their characteristics as well as specific preferences of a user \cite{javed2021review}. Pazzani \cite{pazzani2007content} studies this approach in depth, including how to build user and item profiles. However, as we mentioned before, these models struggle with a limitation to suggest similar papers to other users and they are limited 
by sparse libraries of users. The 
hybrid approach, tries to combine both collaborative and content-based recommendation. It is the most efficient approach which 
includes Collaborative Topic Modelling \cite{wang2011collaborative}, Collaborative Deep Learning \cite{wang2015collaborative} and GRU \cite{10.1145/2959100.2959180}. 

\subsection{Problem formulation} 

We consider a general RS problem where the goal is to model the preference of papers for users by predicting the rate for each user-item pair that the user has not seen or even when the paper is new in the system. Then we make top N recommendations for each user. The two main elements in each RS are users and items: here items are scientific articles and users are researchers. We will use I for users and J for items.\\
We formulate a CF matrix of the user-item rate by $R$. In this matrix, we set the rating for user $i$ for item $j$ as $r_{ij}$ notation. As \cite{10.1145/2959100.2959180} we consider $r_{ij}$ as implicit feedback that is typically defined as follows:

\newcommand{\twopartdef}[4]

\begin{equation}
r_{ij} = \begin{cases}
1, & \text{if $user_i$ interacts with $item_j$ } ,\\
0, & \text{otherwise} 
\end{cases}
\label{eq:1}
\end{equation}
where the $ones$ refer to a positive or observed feedback, and the $zeros$ refer to the missing or unobserved values. The j-th text item is in a sequence
of $n_j$ word tokens, $W_j = (w_1, w_2, w_3,...,w_n)$,  where each token is one of the $V$ words from the vocabulary that we create from our paper collection. \\
For each text item we have number of tags that we represent as follows:

\newcommand{\tagpartdef}[4]

\begin{equation}
t_{jl} = \begin{cases}
1, & \text{if $item_j$ has $tag_l$ } ,\\
0, & \text{otherwise} 
\end{cases}
\label{eq:2}
\end{equation}
Same as with the rate, we assign $t_{jl}$  $one$ if the item $j$ has a tag $l$ otherwise we assign it a $zero$.\\
In this task our goal is to find a list of ranked items that a user has not seen before and make a top N recommendations to them. 
We implement an encoder to map a text input to an embedding vector, learn user, item and tag embedding vectors and make simultaneous predictions of rates and tags. We use $u_i$ for a user with an index $i$, $i_j$ for item with an index $j$ and $t_l$ for tag with an index $l$. In the next section we describe 
our model and how 
to learn efficient text representations from longer sequences of text.

\section{The RHCNN Model}
\begin{figure}[t]
	\centering
	\includegraphics[scale=0.5]{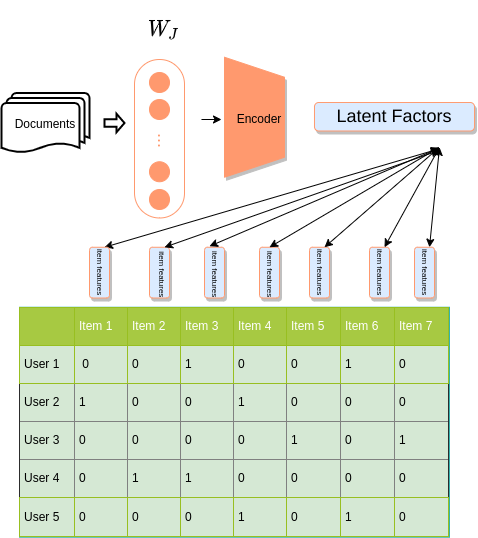}
	\caption{Hybrid CF task for scientific paper Recommendation system}
	\label{fig:CFRS}
\end{figure}
Capturing contextual semantic information from text plays an important role in many applications. 
For this reason several ways to represent structure from text have been designed and implemented \cite{zhang2018deep}. Most of the previous approaches use the bag-of-word approach to represent text but in such case they only rely on word embedding vectors which consider words as independent features. They ignore interactions between words and the order of a text \cite{10.1145/1454008.1454053}. It is evident 
that for long sequences of text an architecture with a capability of capturing 
this semantic information is needed. However, 
using an RNN alone 
is not sufficient to represent the semantics of text as features.  \\
To overcome this issue 
we use RCNN-HW \cite{wen2016learning} as a encoder for mapping text to an embedding vector. This is a 3-layer encoder that captures semantic information from the input and represents it as a vector. This allows us to improve 
the representations for long sequences of text and therefore we are able to represent more fine-grained semantic 
features. 
We expect that this will be able to predict better the meta-data that helps to classify papers by 
capturing long-distance dependencies in texts. 

Figure \ref{fig:CFRS} shows how we add textual features that have been captured by an encoder to the CF task. Our approach for the integration of context into RS is to associate each rating $r_{ij}$ in Equation~\ref{eq:1}, for an item $j$ by a user $i$ with a context description. 
The context adds another dimension to the item-user matrix of the CF. 
A rating $r_{ij}$ is then a tuple of (item, user, context). The context is described as a vector of text mapped to an embedding representation. One drawback of this approach is that it is necessary to compute a similarity of contexts which may be difficult. 
In the next section we describe how our model works.
\subsection{Long sequence text mapping encoder}
Building word and text representations is a basic task in natural language processing. A good representation method can 
learn a good approximation of the grammar and the semantics of natural language. 
Recurrent neural networks (RNNs) can make a use of contextual information when modelling textual representation due to their sequential nature \cite{robinson1987utility}, but long-distance dependencies 
are gradually lost with the gradient propagation \cite{boulanger2012modeling}. RNNs with ``deep'' transition functions remain difficult to train, even when using a gating mechanism. \\ 

We replace the embedding representations of \cite{10.1145/2959100.2959180} with more sophisticated embedding representations that are able to learn from sequential representations of longer documents and can be used for the RS task in a multi-task learning configuration. \\
\begin{figure*}[t]
	\centering
	\includegraphics[scale=0.5]{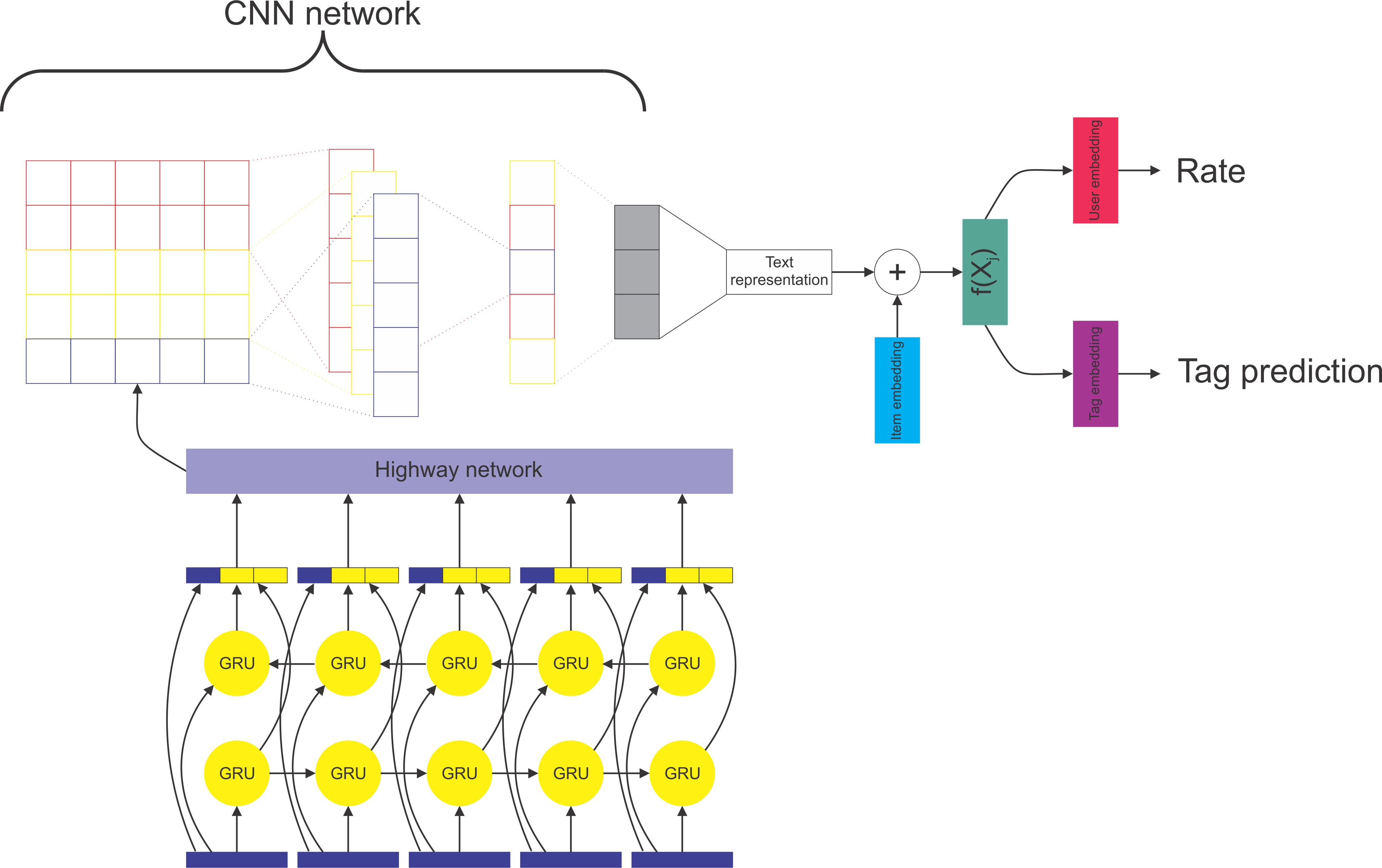}
	\caption{The proposed architecture for multi-task learning. We convert each token to a word embedding vector. This is then fed to a 3-stage network that encodes it to a fixed-length embedding vector. The output of the network are user-paper rates and predicted tags.} 
	\label{fig:rs}
\end{figure*}
\subsubsection{The first stage}
Recurrent neural networks have a potential to map a sequence of a variable length to a fixed-length embedding vector and learn the underlying relations 
from text \cite{tang-etal-2015-document}. 
However, it has been shown that traditional RNNs cannot capture long sequence dependencies because of the vanishing gradient problem \cite{chung2014empirical}. This 
occurs when the back-propagation algorithm attempts to 
updates the weights of units backwards \cite{robinson1987utility}. There are two main approaches to address this issue: Long Short-Term Memory (LSTM) \cite{hochreiter1997long} and Gated Recurrent Unit (GRU) \cite{cho2014properties}. 
Because of a lower complexity and a competitive performance \cite{10.1145/2959100.2959180} we use a (bidirectional)  GRU in this task. 
%
As shown in Figure~\ref{fig:rs} we feed a tokenised sequence as an input to the model. 
We use a pre-training technique to convert a sequence of words to a $k$ dimensional embedding vector and represent the text input as $W_j = (e_1, e_2, e_3,...,e_{nj})$.
The word embeddings are fed to a GRU which produces contextualised word embeddings. 
The contextualised word embeddings are concatenated with the original embeddings which produces the final embeddings represented as a vector of a fixed length.

\subsubsection{The second stage}
Highway layers enable training deep feed-forward networks through the use of adaptive computation.
It is a simple solution to optimise very deep neural networks. This mechanism, like LSTM,  utilises a trainable gating mechanism to adaptively transform or bypass the signal so that the network can 
combine information from structural representations of varying depths \cite{srivastava2015highway}. As 
shown in Equation~\ref{eq:3} a highway has two non-linear gates: 
\begin{equation}
y_t = H(x, W_H).T(x, W_T)+x.C(x, W_c)
\label{eq:3}
\end{equation}

\begin{itemize}
	\item Transformer (T): The $T$ gate controls how much of the activation is passed through 
	\item  Carry (C): the $C$ gate controls how much of the unmodified input is passed through
\end{itemize} 
$H$, $T$ and $C$ typically utilise a sigmoid $(\sigma)$ non-linearity function. In practice for simplicity Equation~\ref{eq:3} can be changed to: 
\begin{equation}
y_t = H(x, W_H).T(x, W_T)+x.(1-T(x, W_T))
\label{eq:4}
\end{equation}

shown in Figure~\ref{fig:rs} the input of this layer is a list of concatenated each word embeddings generated by the GRU. 
Similar to the memory cells in the GRU networks, highway layers allow 
to carry some information from the input directly to the output. In our case the highway is used 
to combine word-level features with document-level features and 
therefore learn semantic representations that require connecting different levels of linguistic structure. 
\\
One advantage of the highway architecture is that the network can learn to dynamically adjust the routing of information based on the current input. 
Also, highway networks allow us to examine how much computation depth is needed for a given problem which can not be easily done with other networks. In this stage we use one highway layer. 
\subsubsection{The third stage}
We add a two layers of CNNs to detect structural patterns of semantic representations in the output of the previous stage and create a fixed-length vector semantic representation of the entire text.
\subsection{Multi-task learning}
\subsubsection{Article recommendation}
The system should recommend to users papers based on their choice history of papers. We use our semantic textual representation to 
provide information about the content of papers.
For this hybrid CF and content-based method we need an item and a user embedding vector as well as item semantic information. 
The rate prediction is now defined as: 
\begin{equation}
\hat{r}_{ij} = bias_i + bias_j + \widetilde u_i ^TF(\widetilde x_j),
\label{eq:5}
\end{equation}  
\begin{equation}
F(\widetilde x_j) = g(x) + \widetilde v_j
\label{eq:6}
\end{equation} 
According to Equation~\ref{eq:6} our system is trained to learn a user $(u_i)$,$(v_j)$ embedding vector, a text embedding vector $g(_x)$ as well as user and item biases. 
Equation~\ref{eq:3} also shows that when an item is new in the system and $v_j$ is unknown, we can still  predict the rate since we have a text embedding vector. \\ 
To learn these parameters we could use any cost function \cite{10.1145/2959100.2959180}. Since we do not want to train unseen and seen data separately we use a weighted mean square: 
\begin{equation}
C_R(\theta)	= \dfrac{1}{|R|} \sum_{(i,j \in{R})} c_{i,j}(\hat{r}_{i,j} - r_{i,j})^2 
\label{eq:7}
\end{equation}

\subsubsection{Meta-data prediction}
Trying to find more information about a user or an item is a priority of each of the RS tasks, 
in our case to successfully predict the rate and the meta-data for items. 
the text embedding vector could potentially be also used for meta-data prediction. This process is useful in cases where an item is new and 
this meta-data can be used for classification. However, the main benefit of this component is in deep training. As user-item rate matrix is highly sparse and some items have been seen very rarely, semantic information from documents will compensate 
such lack of information. \\ 
The tags can be associated with papers by the probability of observing a tag $l$ on an item $j$:
\begin{equation}
t_{ij} = \sigma f(\widetilde{x}_j)^ T t_l
\label{eq:8}
\end{equation}
The embeddings for tags $t_{l}$ has to be learned. For learning the tag embeddings we use a binary cross entropy as a cost function:
\begin{equation}
C_T(\theta)	= \dfrac{1}{|T|} \sum_{j}\sum_{l} t_{j,l} \log { \hat{t}_{j,l} } - (1 - t_{j,l})log(1- \hat{t}_{j,l})
\label{eq:9}
\end{equation} 
\subsection{Multi task learning}
Multi-Task Learning (MTL) is learning several tasks at the same time where knowledge from one task helps another related task as some representations are shared\cite{ruder2017overview}. 
An important motivation of MTL is to alleviate the data sparsity problem where each task has a limited number of labelled data. 
MTL can help reuse existing knowledge and reduce the cost of manual labelling of data. 
MTL models can achieve better performance than single-task counterparts \cite{evgeniou2004regularized} 
because they utilise more data from different learning tasks when compared with single-task learning. 
Common representations are learned in the early layers of a network, while task-specific 
representations are learned in the latter layers \cite{sener2018multi}. 
In our case the tasks are predicting meta-data and user-item rate:
\begin{equation}
C(\theta) = \lambda C_R(\theta)+(1- \lambda)C_T(\theta) 
\label{eq:10}
\end{equation}
where $\lambda$ is a hyper-parameter. 
\section{Datasets and implementation details}
\subsection{Data}
\subsubsection{CiteULike}
We use two datasets of scientific articles and user information for evaluations. CiteULike\footnote{CiteULike \url{http://www.citeulike.org/faq/data.adp}} was a website that offered a service to help researchers create their own library and store, organise, and share the scholarly papers they have read. The rate that our system predicts is based on user-article interaction from this dataset. All datasets come with rich auxiliary information (e.g., title, abstract, citations, and tags) that can be used for the scientific paper recommendation task. \\
First, the CiteULike-a dataset which was collected by \cite{wang2011collaborative} contains information about 5,551 users, 16,980 articles, 204,986 user-article interaction pairs, 46,391 tags, and 44,709 cross-citations between articles. The tags are single-word keywords that are generated by CiteULike researchers when they add an article to their library. The
data sparsity of this dataset is considerably high. Only 0.22\% of the user-article matrix has interactions. \\
Second, the CiteULike-t dataset which was collected by \cite{wang2013collaborative} has 7,947 users, 25,975 articles, 134,860 user-article interaction pairs, 52,946 tags, and 32,565 cross-citations between articles. 
This dataset is sparser than CiteULike-a dataset: here only 0.07\% of the user-article matrix has interactions.

\subsection{Evaluation methodology}
The type of evaluation metrics depends on the type of the recommendation technique. Precision and Recall are the most frequently used evaluation methods. 
A $zero$ rate entry in the $R$ matrix may indicate that a user does not like an article or that the user simply is not aware of it. 
Therefore, this makes Precision not so informative measure. We use Recall instead 
which tell us how successful the retrieval was.
Like most RS we are also using a ranking strategy. The ratings are predicted for items which are then sorted and candidates are selected by returning 
the top M items for each target user. Recall@M per user is defined as:
\begin{equation}
Recall@M = \frac{\text {number of articles user liked in top M}}{\text{total number of articles user liked}}
\end{equation}
The summarised result reported is the average recall over all users. We evaluate our models on a held-out set of articles. 
We consider two recommendation tasks: in-matrix prediction and out-of-matrix prediction. \\
\subsubsection{In matrix prediction (warm start)}
The warm-start prediction is a case where each user has not seen some articles but at least there is one other user that has seen these article. We follow the same procedure as in \cite{10.1145/2959100.2959180, wang2011collaborative} to create the training and testing data. For each user we use a 5-fold cross-validation to split their history between training and testing datasets. We keep papers with less than 5 likes always in the training set. This guarantees that each item has been seen in the training-set at least once. 
Such cases represent 9\% of articles.
\subsubsection{Out of matrix prediction (cold start)}
This case is when we we add completely new papers to the system which have not been previously seen by anyone. For the evaluation model we use a 5-fold cross-validation of papers in the training set and make a test list for users using the folds from the test. Note in this case all users have the same test list and it is guaranteed  that it contains none of the articles in the training set for any user. 
\subsection{Experimental details}
\subsubsection{Text pre-processing}
For each article we concatenate its title and abstract and do not remove the stop words. We use a simple tokeniser. 
We replace numbers with $<$NUM$>$ and those word that appear less than 5 times in the vocabulary with the $<$UNK$>$ token. We use the tf-idf technique to automatically produce tags for some articles that have no tags assigned to them to make a fair comparison for all user-articles pairs. \\

\subsubsection{Training}
We pre-train the GRU in our model on a large corpus of abstracts. We collected 
2 million of abstracts from ACM papers\footnote{\url{https://dl.acm.org}}, Aminer \footnote{\url{https://www.aminer.org}}, ArXiv \footnote{\url{https://arxiv.org}} and CiteULike\footnote{\url{http://www.citeulike.org} [expired]}. 
We compare our model with the GRU \cite{10.1145/2959100.2959180} and Embedding \cite{wang2015collaborative} models. For the Embedding model we compare the results to the results cited in the GRU paper \cite{10.1145/1454008.1454053}. For the GRU we re-implement their model as we do not have access to their code and keep their hyper-parameter settings. For all models we set the word embedding dimensions to $K_w = 200$. 
In the GRU model we follow the setting reported in the paper which is $h_{s1}=400$ and $h_{s2}=200$ for the first and the second RNN hidden state dimensions. For the RHCNN we set $h_{s1}=400$ for the contextualised language model and $h_{s3}=500$ for the convolution layer. 
For Max-pooling to pool the final representation pooling layer we use $p = 200 $ (like $K_w$). Dropout is used on every layer of the network with probabilities of 0.3, 0.3, 0.2 respectively. \\
For both GRU and RHCNN we use the same strategy for training and testing. As \cite{10.1145/1454008.1454053} we do not use different weights for negative and positive samples. As the number of papers with no likes is much larger than papers with likes we use a different strategy for training the model. We create mini-batches of randomly selected users 
and for each user in a mini-batch we sample one positive and one negative item pair. Instead of different weighting for items in the $R$ matrix we set weights as in \cite{10.1145/1454008.1454053} for the cost function in Equation~\ref{eq:7}. 
The models are optimised using a stochastic gradient descent and the Adam algorithm. We run the models in $20k$ randomly sampled data where each mini-batch contains 512 samples. We use early-stopping based Recall on the validation set.

\section{Results}
\begin{center}
	\begin{table*}[ht]
		\centering
		\begin{tabular}{lccc c|| ccc}
			\hline
			& \multicolumn{3}{r}{CiteULike} && \multicolumn{3}{c}{CiteULike} \\
			\cline{3-8} \cline{6-8}  \hline
			\multicolumn{2}{c}{}  & Warm-Start & Cold-Start & Tag-Predictions & Warm-Start & Cold-Start & Tag-Predictions \\\hline
			\multirow{3}{*}{\rotatebox{0}{Models}}  
			& RHCNN & \textbf{42.34} &  \textbf{54.38} &  \textbf{73.34} &  \textbf{49.42} &  \textbf{56.25} & \textbf{77.45} \\\cline{2-8}    
			&GRU & 39.80 & 51.32  & 62.78 &  47.95 &  51.22 & 62.32 \\\cline{2-8}
			& Embed & 36.64 &  47.71 &  60.36 &  43.02 &  38.16& 62.29 \\\cline{2-8}
			\hline  
		\end{tabular}
		\caption{\% Recall@50 for short texts. (higher is better)}
                \label{tbl:sv}

		\begin{tabular}{lccc c|| ccc}
			\hline
			& \multicolumn{3}{r}{CiteULike} && \multicolumn{3}{c}{CiteULike} \\
			\cline{3-8} \cline{6-8}  \hline
			\multicolumn{2}{c}{}  & Warm-Start & Cold-Start & Tag-Predictions & Warm-Start & Cold-Start & Tag-Predictions \\\hline
			\multirow{3}{*}{\rotatebox{0}{Models}}  
			& RHCNN & \textbf{40.55} &  \textbf{54.20} &  \textbf{72.50} &  \textbf{49.22} &  50.36 & \textbf{70.80} \\\cline{2-8}    
			&GRU & 37.90 & 50.24  & 58.45 &  44.68 &  42.68 & 59.43 \\\cline{2-8}
			\hline  
		\end{tabular}
		\caption{\% Recall@50 for long texts. (higher is better)}
                \label{tbl:lv}
	\end{table*}
\end{center}

In this section we discuss the results and a comparison of the RHCNN models with the GRU \cite{10.1145/1454008.1454053} and the Embedding \cite{wang2015collaborative} model. 
All of the models have been implemented in the hybrid MTL configuration. We consider two experiments, one containing texts of $200$ tokens (short version) and another containing texts of $400$ tokens (long version). For the long experiment we only compare RHCNN with GRU since they are mapping sequences of text to embedding vectors and the Embedding model only uses word embedding vectors. In this context the order or the number of tokens is not important. 
Our main goal is to have a model that performs over longer sequences. We set $Top@M$ to $M = 10, 20, ..., 100$. Table~\ref{tbl:sv} and \ref{tbl:lv} summarise Recall@50 for all models in both long and short versions.
\subsubsection{Warm-Start}
Figures \ref{fig:ca-w} and \ref{fig:ct-w} show the results for the RHCNN, GRU, and the Embedding model using both CiteULike-a and CiteULike-t datasets using the short version of the text input. 
RHCNN works well on both datasets. However when M becomes large fewer user ratings are available to help CF give good recommendations and the contribution of the content becomes more important. Considering the RHCNN performance, even in the warm-start condition the role of the contextualised semantic representations is very important. 
RHCNN recommends articles $8.7\%$ on CiteULike-a and $8.5\%$ on CiteULike-t better than other models with recall@50.\\
It is therefore important that encoder can extract semantically information from texts that are longer. 
Longer sequences mean there may be more semantic information but also note that it is hard for neural architecture to find dependencies between such information.
Figure~\ref{fig:ca-wl} and \ref{fig:ct-wl} show results comparing RHCNN with GRU. We see 
that over longer sequences our model can make better predictions than GRU as it can better relate the semantic information in the text.

\subsubsection{Cold-Start}
Cold-start recommendations are made on new articles, unseen during training. 
Therefore, there is no embedding vector for an item and $v_j$ in Equation~\ref{eq:6} is unknown so we only rely on the item's content. In Figure~\ref{fig:ca-c} and \ref{fig:ct-c} we see that RHCNN performs considerably better than either GRU and Embedding over short texts. 
On CiteULike-a RHCNN performs better by $10\%$ and on CiteULike-t by $13\%$ than other models. In \cite{10.1145/1454008.1454053} Trapit et al. report the performance on CiteULike-t higher than on CiteULike-a attributing this to more sparsity in this dataset and because results in this dataset depend on contextual information. 
In Figure~\ref{fig:ca-cl} and \ref{fig:ct-cl} we see our advantage when sequences of text are long.\\
Overall, we can conclude that our model is flexible over different conditions. We see that our model performs better over both long and short texts and in cold and warm-start conditions than other models.


\begin{center}
	\begin{figure}[t]
		\caption{CiteULike-a Warm-Start, short texts}\label{fig:ca-w}
		\begin{tikzpicture}
		\begin{axis}[
		title={},
		ylabel={recall},
		xmin=0, xmax=100,
		ymin=0, ymax=0.8,
		xtick={0,20,40,60,80,100},
		ytick={0.0, 0.10, 0.20, 0.30, 0.40, 0.50, 0.60, 0.70},
		legend pos=north west,
		ymajorgrids=true,
		grid style=dashed,
		]
		\addplot[
		thick,
		color=blue,
		mark=square,
		]
		coordinates {
			(10,0.21)(20,0.30)(30,0.35)(40,0.39)(50,0.42)(60,0.45)(70, 0.48)(80, 0.50)(90, 0.53)(100, 0.55)
		};
		\addplot[
		thick,
		color=red,
		mark=triangle,
		]
		coordinates {
			(10,0.17)(20,0.23)(30,0.31)(40,0.35)(50,0.38)(60,0.41)(70, 0.45)(80, 0.47)(90, 0.49)(100, 0.51)
		};
		\addplot[
		thick,
		color=green,
		mark=star,
		]
		coordinates {
			(10,0.168)(20,0.23)(30,0.30)(40,0.33)(50,0.36)(60,0.38)(70, 0.41)(80, 0.45)(90, 0.47)(100, 0.49)
		};
		\legend{RHCNN, GRU, Embedding}
		\end{axis}
		
		\end{tikzpicture}
	\end{figure}

	\begin{figure}[t]
		\caption{CiteULike-t Warm-Start, short texts}\label{fig:ct-w}
		\begin{tikzpicture}
		\begin{axis}[
		title={},
		ylabel={recall},
		xmin=0, xmax=100,
		ymin=0, ymax=0.8,
		xtick={0,20,40,60,80,100},
		ytick={0.0, 0.10, 0.20, 0.30, 0.40, 0.50, 0.60, 0.70},
		legend pos=north west,
		ymajorgrids=true,
		grid style=dashed,
		]
		\addplot[
		thick,
		color=blue,
		mark=square,
		]
		coordinates {
			(10,0.26)(20,0.35)(30,0.41)(40,0.45)(50,0.49)(60,0.52)(70, 0.55)(80, 0.57)(90, 0.60)(100, 0.62)
		};
		\addplot[
		thick,
		color=red,
		mark=triangle,
		]
		coordinates {
			(10,0.22)(20,0.33)(30,0.39)(40,0.42)(50,0.47)(60,0.50)(70, 0.52)(80, 0.54)(90, 0.56)(100, 0.59)
		};
		\addplot[
		thick,
		color=green,
		mark=star,
		]
		coordinates {
			(10,0.20)(20,0.30)(30,0.37)(40,0.39)(50,0.43)(60,0.47)(70, 0.49)(80, 0.51)(90, 0.53)(100, 0.55)
		};
		\legend{RHCNN, GRU, Embedding}
		\end{axis}
		
		\end{tikzpicture}
              \end{figure}

\begin{center}
	\begin{figure}[t]
		\caption{CiteULike-a Warm-Start, long texts}\label{fig:ca-wl}
		\begin{tikzpicture}
		\begin{axis}[
		title={},
		ylabel={recall},
		xmin=0, xmax=100,
		ymin=0, ymax=0.8,
		xtick={0,20,40,60,80,100},
		ytick={0.0, 0.10, 0.20, 0.30, 0.40, 0.50, 0.60, 0.70},
		legend pos=north west,
		ymajorgrids=true,
		grid style=dashed,
		]
		\addplot[
		thick,
		color=blue,
		mark=square,
		]
		coordinates {
			(10,0.19)(20,0.27)(30,0.33)(40,0.37)(50,0.40)(60,0.43)(70, 0.46)(80, 0.48)(90, 0.50)(100, 0.52)
		};
		\addplot[
		thick,
		color=red,
		mark=triangle,
		]
		coordinates {
			(10,0.17)(20,0.23)(30,0.30)(40,0.34)(50,0.37)(60,0.40)(70, 0.42)(80, 0.45)(90, 0.46)(100, 0.49)
		};
		\legend{RHCNN, GRU}
		\end{axis}
		
		\end{tikzpicture}
	\end{figure}
	
	\begin{figure}[t]
		\caption{CiteULike-t Warm-Start, long texts}\label{fig:ct-wl}
		\begin{tikzpicture}
		\begin{axis}[
		title={},
		ylabel={recall},
		xmin=0, xmax=100,
		ymin=0, ymax=0.8,
		xtick={0,20,40,60,80,100},
		ytick={0.0, 0.10, 0.20, 0.30, 0.40, 0.50, 0.60, 0.70},
		legend pos=north west,
		ymajorgrids=true,
		grid style=dashed,
		]
		\addplot[
		thick,
		color=blue,
		mark=square,
		]
		coordinates {
			(10,0.26)(20,0.35)(30,0.41)(40,0.45)(50,0.49)(60,0.52)(70, 0.55)(80, 0.57)(90, 0.59)(100, 0.60)
		};
		\addplot[
		thick,
		color=red,
		mark=triangle,
		]
		coordinates {
			(10,0.22)(20,0.30)(30,0.36)(40,0.40)(50,0.44)(60,0.47)(70, 0.49)(80, 0.51)(90, 0.53)(100, 0.56)
		};
		\legend{RHCNN, GRU}
		\end{axis}
		
		\end{tikzpicture}
		
	\end{figure}


	\begin{figure}[t]
		\caption{CiteULike-a Cold-Start, short texts}\label{fig:ca-c}
		\begin{tikzpicture}
		\begin{axis}[
		title={},
		ylabel={recall},
		xmin=0, xmax=100,
		ymin=0, ymax=0.8,
		xtick={0,20,40,60,80,100},
		ytick={0.0, 0.10, 0.20, 0.30, 0.40, 0.50, 0.60, 0.70},
		legend pos=north west,
		ymajorgrids=true,
		grid style=dashed,
		]
		\addplot[
		thick,
		color=blue,
		mark=square,
		]
		coordinates {
			(10,0.27)(20,0.36)(30,0.45)(40,0.51)(50,0.54)(60,0.57)(70, 0.60)(80, 0.62)(90, 0.65)(100, 0.67)
		};
		\addplot[
		thick,
		color=red,
		mark=triangle,
		]
		coordinates {
			(10,0.22)(20,0.32)(30,0.40)(40,0.45)(50,0.50)(60,0.53)(70, 0.55)(80, 0.57)(90, 0.60)(100, 0.62)
		};
		\addplot[
		thick,
		color=green,
		mark=star,
		]
		coordinates {
			(10,0.16)(20,0.23)(30,0.30)(40,0.35)(50,0.42)(60,0.45)(70, 0.47)(80, 0.50)(90, 0.53)(100, 0.55)
		};
		\legend{RHCNN, GRU, Embedding}
		\end{axis}
		
		\end{tikzpicture}
	\end{figure}
		
		\begin{figure}[t]
			\caption{CiteULike-t Cold-Start, short texts}\label{fig:ct-c}
		\begin{tikzpicture}
		\begin{axis}[
		title={},
		ylabel={recall},
		xmin=0, xmax=100,
		ymin=0, ymax=0.8,
		xtick={0,20,40,60,80,100},
		ytick={0.0, 0.10, 0.20, 0.30, 0.40, 0.50, 0.60, 0.70},
		legend pos=north west,
		ymajorgrids=true,
		grid style=dashed,
		]
		\addplot[
		thick,
		color=blue,
		mark=square,
		]
		coordinates {
			(10,0.27)(20,0.42)(30,0.47)(40,0.51)(50,0.56)(60,0.59)(70, 0.62)(80, 0.64)(90, 0.66)(100, 0.69)
		};
		\addplot[
		thick,
		color=red,
		mark=triangle,
		]
		coordinates {
			(10,0.23)(20,0.36)(30,0.41)(40,0.45)(50,0.52)(60,0.55)(70, 0.57)(80, 0.60)(90, 0.65)(100, 0.66)
		};
		\addplot[
		thick,
		color=green,
		mark=star,
		]
		coordinates {
			(10,0.10)(20,0.20)(30,0.25)(40,0.33)(50,0.39)(60,0.42)(70, 0.45)(80, 0.50)(90, 0.53)(100, 0.55)
		};
		\legend{RHCNN, GRU, Embedding}
		\end{axis}
		
		\end{tikzpicture}
	
	\end{figure}
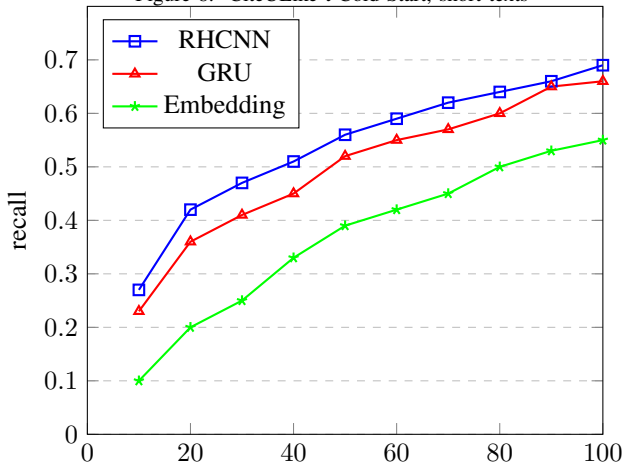
\end{center}


	\begin{figure}[t]
		\caption{CiteULike-a Cold-Start, long texts}\label{fig:ca-cl}
		\begin{tikzpicture}
		\begin{axis}[
		title={},
		ylabel={recall},
		xmin=0, xmax=100,
		ymin=0, ymax=0.8,
		xtick={0,20,40,60,80,100},
		ytick={0.0, 0.10, 0.20, 0.30, 0.40, 0.50, 0.60, 0.70},
		legend pos=north west,
		ymajorgrids=true,
		grid style=dashed,
		]
		\addplot[
		thick,
		color=blue,
		mark=square,
		]
		coordinates {
			(10,0.27)(20,0.36)(30,0.45)(40,0.51)(50,0.54)(60,0.57)(70, 0.60)(80, 0.62)(90, 0.65)(100, 0.67)
		};
		\addplot[
		thick,
		color=red,
		mark=triangle,
		]
		coordinates {
			(10,0.22)(20,0.32)(30,0.40)(40,0.45)(50,0.50)(60,0.53)(70, 0.55)(80, 0.57)(90, 0.60)(100, 0.62)
		};
		\legend{RHCNN, GRU}
		\end{axis}
		
		\end{tikzpicture}
	\end{figure}
	
	\begin{figure}[t]
		\caption{CiteULike-t Cold-Start, long texts}\label{fig:ct-cl}
		\begin{tikzpicture}
		\begin{axis}[
		title={},
		ylabel={recall},
		xmin=0, xmax=100,
		ymin=0, ymax=0.8,
		xtick={0,20,40,60,80,100},
		ytick={0.0, 0.10, 0.20, 0.30, 0.40, 0.50, 0.60, 0.70},
		legend pos=north west,
		ymajorgrids=true,
		grid style=dashed,
		]
		\addplot[
		thick,
		color=blue,
		mark=square,
		]
		coordinates {
			(10,0.27)(20,0.35)(30,0.41)(40,0.45)(50,0.50)(60,0.53)(70, 0.56)(80, 0.59)(90, 0.62)(100, 0.65)
		};
		\addplot[
		thick,
		color=red,
		mark=triangle,
		]
		coordinates {
			(10,0.19)(20,0.25)(30,0.32)(40,0.38)(50,0.42)(60,0.46)(70, 0.51)(80, 0.54)(90, 0.58)(100, 0.61)
		};
		\legend{RHCNN, GRU}
		\end{axis}
		
		\end{tikzpicture}
		
	\end{figure}
\end{center}

\subsubsection{Tag Predictions}
Predicting the meta-data such tags for articles is expected to help the RS in predicting user-item rates. This is because tags encode topics that might be taken into account when determining the preference. We report recall@50  for tag predictions on articles in Table~\ref{tbl:sv} and \ref{tbl:lv}. In the warm-start scenario we train articles on their tags. Hence, in this task tags prediction is meaningless. 
Therefore, we learn tag prediction only in the cold-start scenario. The results indicate that tag prediction is much better in the RHCNN model than other models and we attribute this to the benefits of contextualised semantic information learned with Highway Networks. On CiteULike-a with recall@50 and with $400$ tokens we get a $\%14$ increase in performance  and on CiteULike-t $\% 10.45$ compared with the GRU model.
\section{Conclusions and future work}
We proposed a novel method for recommendation systems of scientific papers but which could also be extended to other recommendation tasks. 
We demonstrate that analysing the paper's content and extracting unseen semantic information from text is one of the strengths in creating a hybrid recommendation system that relies both on collaborative filtering and item features. We employed a neural network that uses a highway mechanism to 
relate semantic information that can be presented as different long or short dependencies over larger sequences of text. 
Our results indicate that the proposed system can deal with the difficulty of a cold start and user-item matrix sparsity. 
For the next step, we plan to experiment with different sampling methods to reduce the amount of data needed to train the network without effecting its performance as well as compare our semantic representations that were tailor-trained on the scientific texts with the general semantic representations obtained through BERT.


\bibliographystyle{plain}
\bibliography{bibliography}

\begin{thebibliography}{10}

\bibitem{bai2019scientific}
Xiaomei Bai, Mengyang Wang, Ivan Lee, Zhuo Yang, Xiangjie Kong, and Feng Xia.
\newblock Scientific paper recommendation: A survey.
\newblock {\em Ieee Access}, 7:9324--9339, 2019.

\bibitem{10.1145/2959100.2959180}
Trapit Bansal, David Belanger, and Andrew McCallum.
\newblock Ask the gru: Multi-task learning for deep text recommendations.
\newblock RecSys '16, page 107–114, New York, NY, USA, 2016. Association for
  Computing Machinery.

\bibitem{10.1145/1454008.1454053}
Toine Bogers and Antal van~den Bosch.
\newblock Recommending scientific articles using citeulike.
\newblock RecSys '08, New York, NY, USA, 2008. Association for Computing
  Machinery.

\bibitem{boulanger2012modeling}
Nicolas Boulanger-Lewandowski, Yoshua Bengio, and Pascal Vincent.
\newblock Modeling temporal dependencies in high-dimensional sequences:
  Application to polyphonic music generation and transcription.
\newblock {\em arXiv preprint arXiv:1206.6392}, 2012.

\bibitem{bulut2018paper}
Bet{\"u}l Bulut, Buket Kaya, Reda Alhajj, and Mehmet Kaya.
\newblock A paper recommendation system based on user's research interests.
\newblock In {\em 2018 IEEE/ACM International Conference on Advances in Social
  Networks Analysis and Mining (ASONAM)}, pages 911--915. IEEE, 2018.

\bibitem{cho2014properties}
Kyunghyun Cho, Bart van Merrienboer, Dzmitry Bahdanau, and Yoshua Bengio.
\newblock On the properties of neural machine translation: Encoder-decoder
  approaches, 2014.

\bibitem{cho2014learning}
Kyunghyun Cho, Bart Van~Merri{\"e}nboer, Caglar Gulcehre, Dzmitry Bahdanau,
  Fethi Bougares, Holger Schwenk, and Yoshua Bengio.
\newblock Learning phrase representations using rnn encoder-decoder for
  statistical machine translation.
\newblock {\em arXiv preprint arXiv:1406.1078}, 2014.

\bibitem{chung2014empirical}
Junyoung Chung, Caglar Gulcehre, KyungHyun Cho, and Yoshua Bengio.
\newblock Empirical evaluation of gated recurrent neural networks on sequence
  modeling.
\newblock {\em arXiv preprint arXiv:1412.3555}, 2014.

\bibitem{dai2018low}
Tao Dai, Tianyu Gao, Li~Zhu, Xiaoyan Cai, and Shirui Pan.
\newblock Low-rank and sparse matrix factorization for scientific paper
  recommendation in heterogeneous network.
\newblock {\em IEEE Access}, 6:59015--59030, 2018.

\bibitem{DHANDA2016483}
Mahak Dhanda and Vijay Verma.
\newblock Recommender system for academic literature with incremental dataset.
\newblock {\em Procedia Computer Science}, 89:483--491, 2016.

\bibitem{evgeniou2004regularized}
Theodoros Evgeniou and Massimiliano Pontil.
\newblock Regularized multi--task learning.
\newblock In {\em Proceedings of the tenth ACM SIGKDD international conference
  on Knowledge discovery and data mining}, pages 109--117, 2004.

\bibitem{ghazanfar2010scalable}
Mustansar~Ali Ghazanfar and Adam Prugel-Bennett.
\newblock A scalable, accurate hybrid recommender system.
\newblock In {\em 2010 Third International Conference on Knowledge Discovery
  and Data Mining}, pages 94--98. IEEE, 2010.

\bibitem{hassan2017personalized}
Hebatallah A~Mohamed Hassan.
\newblock Personalized research paper recommendation using deep learning.
\newblock In {\em Proceedings of the 25th conference on user modeling,
  adaptation and personalization}, pages 327--330, 2017.

\bibitem{herlocker2002empirical}
Jon Herlocker, Joseph~A Konstan, and John Riedl.
\newblock An empirical analysis of design choices in neighborhood-based
  collaborative filtering algorithms.
\newblock {\em Information retrieval}, 5(4):287--310, 2002.

\bibitem{hochreiter1997long}
Sepp Hochreiter and J{\"u}rgen Schmidhuber.
\newblock Long short-term memory.
\newblock {\em Neural computation}, 9(8):1735--1780, 1997.

\bibitem{javed2021review}
Umair Javed, Kamran Shaukat, Ibrahim~A Hameed, Farhat Iqbal, Talha~Mahboob
  Alam, and Suhuai Luo.
\newblock A review of content-based and context-based recommendation systems.
\newblock {\em International Journal of Emerging Technologies in Learning
  (iJET)}, 16(3):274--306, 2021.

\bibitem{koren2015advances}
Yehuda Koren and Robert Bell.
\newblock Advances in collaborative filtering.
\newblock {\em Recommender systems handbook}, pages 77--118, 2015.

\bibitem{liu2015car}
Haifeng Liu, Zhuo Yang, Ivan Lee, Zhenzhen Xu, Shuo Yu, and Feng Xia.
\newblock Car: Incorporating filtered citation relations for scientific article
  recommendation.
\newblock In {\em 2015 IEEE International Conference on Smart
  City/SocialCom/SustainCom (SmartCity)}, pages 513--518. IEEE, 2015.

\bibitem{LU201512}
Jie Lu, Dianshuang Wu, Mingsong Mao, Wei Wang, and Guangquan Zhang.
\newblock Recommender system application developments: A survey.
\newblock {\em Decision Support Systems}, 74:12--32, 2015.

\bibitem{pazzani2007content}
Michael~J Pazzani and Daniel Billsus.
\newblock Content-based recommendation systems.
\newblock In {\em The adaptive web}, pages 325--341. Springer, 2007.

\bibitem{pouyanfar2018survey}
Samira Pouyanfar, Saad Sadiq, Yilin Yan, Haiman Tian, Yudong Tao, Maria~Presa
  Reyes, Mei-Ling Shyu, Shu-Ching Chen, and Sundaraja~S Iyengar.
\newblock A survey on deep learning: Algorithms, techniques, and applications.
\newblock {\em ACM Computing Surveys (CSUR)}, 51(5):1--36, 2018.

\bibitem{robinson1987utility}
AJ~Robinson and Frank Fallside.
\newblock {\em The utility driven dynamic error propagation network}.
\newblock University of Cambridge Department of Engineering Cambridge, 1987.

\bibitem{ruder2017overview}
Sebastian Ruder.
\newblock An overview of multi-task learning in deep neural networks.
\newblock {\em arXiv preprint arXiv:1706.05098}, 2017.

\bibitem{sener2018multi}
Ozan Sener and Vladlen Koltun.
\newblock Multi-task learning as multi-objective optimization.
\newblock {\em arXiv preprint arXiv:1810.04650}, 2018.

\bibitem{sharma2013survey}
Lalita Sharma and Anju Gera.
\newblock A survey of recommendation system: Research challenges.
\newblock {\em International Journal of Engineering Trends and Technology
  (IJETT)}, 4(5):1989--1992, 2013.

\bibitem{srivastava2015highway}
Rupesh~Kumar Srivastava, Klaus Greff, and J{\"u}rgen Schmidhuber.
\newblock Highway networks.
\newblock {\em arXiv preprint arXiv:1505.00387}, 2015.

\bibitem{su2009survey}
Xiaoyuan Su and Taghi~M Khoshgoftaar.
\newblock A survey of collaborative filtering techniques.
\newblock {\em Advances in artificial intelligence}, 2009, 2009.

\bibitem{sun2014leveraging}
Jianshan Sun, Jian Ma, Zhiying Liu, and Yajun Miao.
\newblock Leveraging content and connections for scientific article
  recommendation in social computing contexts.
\newblock {\em The Computer Journal}, 57(9):1331--1342, 2014.

\bibitem{tang-etal-2015-document}
Duyu Tang, Bing Qin, and Ting Liu.
\newblock Document modeling with gated recurrent neural network for sentiment
  classification.
\newblock In {\em Proceedings of the 2015 Conference on Empirical Methods in
  Natural Language Processing}, pages 1422--1432, Lisbon, Portugal, September
  2015. Association for Computational Linguistics.

\bibitem{wang2011collaborative}
Chong Wang and David~M Blei.
\newblock Collaborative topic modeling for recommending scientific articles.
\newblock In {\em Proceedings of the 17th ACM SIGKDD international conference
  on Knowledge discovery and data mining}, pages 448--456, 2011.

\bibitem{wang2013collaborative}
Hao Wang, Binyi Chen, and Wu-Jun Li.
\newblock Collaborative topic regression with social regularization for tag
  recommendation.
\newblock In {\em Twenty-Third International Joint Conference on Artificial
  Intelligence}, 2013.

\bibitem{wang2015collaborative}
Hao Wang, Naiyan Wang, and Dit-Yan Yeung.
\newblock Collaborative deep learning for recommender systems.
\newblock In {\em Proceedings of the 21th ACM SIGKDD international conference
  on knowledge discovery and data mining}, pages 1235--1244, 2015.

\bibitem{wen2016learning}
Ying Wen, Weinan Zhang, Rui Luo, and Jun Wang.
\newblock Learning text representation using recurrent convolutional neural
  network with highway layers.
\newblock {\em arXiv preprint arXiv:1606.06905}, 2016.

\bibitem{west2016recommendation}
Jevin~D West, Ian Wesley-Smith, and Carl~T Bergstrom.
\newblock A recommendation system based on hierarchical clustering of an
  article-level citation network.
\newblock {\em IEEE Transactions on Big Data}, 2(2):113--123, 2016.

\bibitem{zarrinkalam2013semcir}
Fattane Zarrinkalam and Mohsen Kahani.
\newblock Semcir: A citation recommendation system based on a novel semantic
  distance measure.
\newblock {\em Program}, 2013.

\bibitem{zhang2018deep}
Lei Zhang, Shuai Wang, and Bing Liu.
\newblock Deep learning for sentiment analysis: A survey.
\newblock {\em Wiley Interdisciplinary Reviews: Data Mining and Knowledge
  Discovery}, 8(4):e1253, 2018.

\end{thebibliography}

\end{document}